\documentclass[a4paper,11pt]{article}
\usepackage[utf8x]{inputenc}
\usepackage{amssymb,amsfonts,amstext,amsmath,graphicx,epic,amsthm,color,times}

\title{Linear stability  analysis of  fluid flow  between two   parallel 
porous stationary   plates  with  small suction and  injection}
\author{L. A. Hinvi\footnote{ahinvi@yahoo.fr}\hspace{0.08cm}, A. V. Monwanou\footnote{movins$2008$@yahoo.fr} \hspace{0.05cm}  and J. B. Chabi
   Orou\footnote{Author to whom correspondence should be addressed: jchabi@yahoo.fr}}

\begin{document}
\maketitle{Institut de Math\'ematiques et de Sciences Physiques, BP: 613 Porto Novo, B\'enin}
\begin{abstract}
In this work,  the  linear stability  of the viscous incompressible
fluid flow between two parallel horizontal porous stationary  plates 
with the assumption that there is a small constant   suction at upper plate and a small constant  injection at the
lower plate is studied.The Navier-Stokes and  continuous equations  
are reduced to  an equation  modified by the   suction  Reynolds number,
which we call modified Orr-Sommerfeld equation. 
This equation is rewritten as an eigenvalue problem and  is  solved
numerically using  Matlab (Windows Version).
The  effect of small suction 
Reynolds number  on the linear stability fluid flow is discussed.
 
\end{abstract}
{\bf{Keywords:}} Porous  parallel plates, linear stability,  
small suction  Reynolds number, modified Orr-Sommerfeld equation.

\section{Introduction}
The flow through porous boundaries is of great importance both in technological as well as biophysical flows. Examples
of this are found in soil mechanics, the aviation industry, transpiration cooling,
food preservation, cosmetic industry, blood flow and artificial
dialysis. A large number of theoretical investigations dealing
with steady incompressible laminar flow with either injection
or suction at the boundaries (\cite{1} -\cite{2}) have appeared during the last few
decades. Several authors \cite{3} have studied the steady
laminar flow of an incompressible viscous fluid in a two-
dimensional channel with parallel porous walls.
 Rioual and al. $(1996)$ studied the power balance of a flat plate used 
as an airplane wing and found that there is an optimum speed of suction 
that reduces drag and weaken energy consumption.
 Recent experimental studies show that suction damps growth of 
 disturbances induced by free stream turbulence and transition is delayed/prevented 
 (Fransson and Alfredsson $2003$).
Aspiration may also be used as a tool to induce transition, instead of the delay,
if for example the wall material chosen is not able to provide continuous suction 
(MacManus and Eaton, $2000$) or if it is applied non-uniformly
(Roberts and Floryan, $2001$) \cite{4}.

In this work,  the flow of the incompressible viscous fluid between 
two parallel horizontal stationary porous plates  with  the assumption that there is
constant small  suction at upper plate and a constant small injection at the
lower plate is  considered. Speeds suction and injection are assumed  uniform and same standard 
 $ V^{*}_{\omega}$ (see figure (\ref{fig:1})) below  \cite{3}. The objective  is to study the effect 
of number $R_{e\omega}$ of  extradvective term introduced in an
Orr-Sommerfeld  equation  by small  suction velocity on  the stability of  the fluid flow. 
Such attempt has been made earlier 
by  E. Niklas Davidsson  and L. Hakan Gustavsson the $2000s$ ( \cite{1} and \cite{4})
the case of the   boundary layers (a flat plate-law)  with uniform wall suction by
normalizing all the components of the velocity  with the free stream velocity $U^{*}$.
 Before E. Niklas Davidsson and al. ; V. M. Soundalgekar, V. G. Divekar and al. ($1968-1972$) 
studied the flow quilt with suction at  the stationary plate and  laminar slip-flow through 
uniform circular pipe with small suction ( \cite{5}-\cite{6}). 
All these studies confirm the stabilizing effect of small  wall suction on a flow.

The paper is organized as follows.
In the second section the modified Orr-Sommerfeld equation governing the 
stability analysis in the  Poiseuille  two parallel horizontal 
porous stationary  plates flow will be checked. In the third
section  an  analysis of  the small suction Reynolds number   $R_{e\omega}$ 
stabilizing effect will be investigated. The conclusions will be presented in the
final section.

\section{modified Orr-Sommerfeld equation}

The motion of a fluid is governed by the conservation laws, such as 
conservation of mass and momentum. The momentum of the fluid changes
due to the forces acting upon it, and is governed by Newton’s second
law of motion.
Fluids at low velocity compared to the speed of sound can be 
considered incompressible, i.e. the density of the fluid is constant.
The stress acting on a fluid element may be linearly proportional
to the local pressure and velocity gradient.
For an incompressible, newtonian fluid the momentum and conservation  equations
yield
\begin{eqnarray}\label{eqa1}
 \frac{\partial{\tilde{u}_{i}^{*}}}{\partial{\tilde{t}^{*}}} +
 \tilde{u}_{j}^{*}\frac{\partial{\tilde{u}_{i}^{*}}}{\partial{\tilde{x}_{j}^{*}}} &=&
 -\frac{1}{\rho}\frac{\partial{\tilde{p}}^{*}}{\partial{\tilde{x}_{i}^{*}}}+
 {\nu}\frac{\partial^{2}{\tilde{u}_{i}^{*}}}{{\partial{\tilde{x}_{j}^{*}}}^{2}}\\ \label{eqa2}
 \frac{\partial{\tilde{u}_{i}^{*}}}{\partial{\tilde{x}_{i}^{*}}}&=&0 
 \end{eqnarray}
when written by Einsteins' summation convention. These are the 
Navier-Stokes (N$-$S) equations. Here
$\tilde{x}^{*}_{i}$ and  $\tilde{u}^{*}_{i}$  are the  $i^{ieme}$ component of
the space and velocity vectors, as commonly in the literature denoted
$(\tilde{x}^{*},  \tilde{y}^{*}, \tilde{z}^{*})$ and  $(\tilde{u}^{*}, \tilde{v}^{*},
 \tilde{w}^{*})$ in the streamwise, wall-normal and spanwise directions, respectively.
Also, $\tilde{P}^{*}$ is the pressure, $\rho$ is the density and $\nu$
is the kinematical viscosity of the fluid.

\begin{figure}[htbp]
 \begin{center}
 \includegraphics[width=7cm]{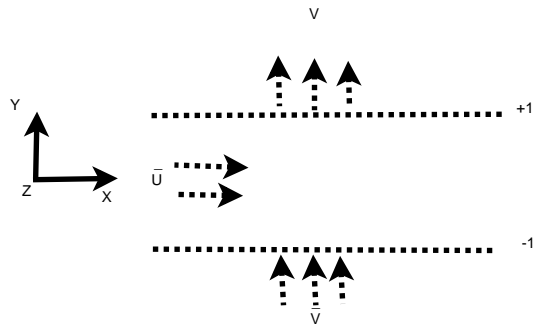} 
 \end{center}
 \caption{Poiseuille porous parallel flat plates flow.} 
 \label{fig:1}
 \end{figure}

 We Considered  the viscous and incompressible fluid flow  between two parallel  horizontal stationary porous flat plates     
 under a constant gradient of pressure with small wall  suction and injection
in  the figure (\ref{fig:1}) below \cite{3}.
\newpage
 The coordinates (streamwise $x$, the wall-normal $y$ and spanwise $z$) are scaled
 with the length scale $h^{*}$ ( distance between the two walls). The streamwise
 and spanwise velocities $u$ and $w$, respectively, are scaled with the  streamwise 
 free stream velocity $U^{*}$ while the wall-normal
velocity $v$ is  scaled with the characteristic velocity of suction and injection 
$V_{\omega}^{*}$. The pressure $p$ is scaled with
$\rho U^{*2}$ and the time $t$ with $\frac{h^{*}}{U^{*}}$.
The Reynolds' numbers  used here are defined as $R_{e}=\frac{U^{*}h^{*}}{\nu}$ 
and  suction Reynolds' number $R_{e\omega}=\frac{V_{\omega}^{*}h^{*}}{\nu}$ where $\nu$ 
is kinematic viscosity.\\
We want to study the linear stability of a high Reynolds  $R_{e}$  and
small suction  Renynolds $R_{e\omega}$ ($0\leq R_{e\omega} \leq 1$)  numbers flow. 
The non-dimensional Navier-Stokes  and continuous (Conservation of mass) linearized 
equations (\ref{eqa1}) and (\ref{eqa2}) 
 where  the variables are normalized as above take the following forms

\begin{eqnarray}
   \frac{\partial{{\tilde{u}}}}{\partial{{\tilde{t}}}} +
   {\tilde{u}}\frac{\partial{\tilde{u}}}{\partial{\tilde{x}}}+ 
   \frac{R_{e \omega}}{R_{e}} {\tilde{v}}\frac{\partial{\tilde{u}}}{\partial{\tilde{y}}} +
  {\tilde{w}}\frac{\partial{\tilde{u}}}{\partial{\tilde{z}}}  
  &=&  -\frac{\partial{\tilde{p}}}{\partial{\tilde{x}}}+
  \frac{1}{R_{e}} {\bigtriangledown}^{2}{\tilde{u}},\label{eq1}\\
  \frac{\partial{{\tilde{v}}}}{\partial{{\tilde{t}}}} +
  {\tilde{u}}\frac{\partial{\tilde{v}}}{\partial{\tilde{x}}}+
  \frac{R_{e \omega}}{R_{e}} {\tilde{v}}\frac{\partial{\tilde{v}}}{\partial{\tilde{y}}} +
  {\tilde{w}}\frac{\partial{\tilde{v}}}{\partial{\tilde{z}}} 
  &=&  -\frac{R_{e}}{R_{e \omega}} \frac{\partial{\tilde{p}}}{\partial{\tilde{y}}}+ 
  \frac{1}{R_{e}} {\bigtriangledown}^{2}{\tilde{v}},\label{eq2}\\
\frac{\partial{{\tilde{w}}}}{\partial{{\tilde{t}}}} + 
{\tilde{u}}\frac{\partial{\tilde{w}}}{\partial{\tilde{x}}}+ 
\frac{R_{e \omega}}{R_{e}}{\tilde{v}}\frac{\partial{\tilde{w}}}{\partial{\tilde{y}}} +
  {\tilde{w}}\frac{\partial{\tilde{w}}}{\partial{\tilde{z}}}  
  &=& -\frac{\partial{\tilde{p}}}{\partial{\tilde{z}}}+ 
  \frac{1}{R_{e}} {\bigtriangledown}^{2}{\tilde{w}}.\label{eq3}
  \end{eqnarray}
\begin{eqnarray}
  \frac{\partial{\tilde{u}}}{\partial{\tilde{x}}} + 
  \frac{R_{e \omega}}{R_{e}}\frac{\partial{\tilde{v}}}{\partial{\tilde{y}}} +
 \frac{\partial{\tilde{w}}}{\partial{\tilde{z}}} &=&0, \label{eq4}
\end{eqnarray}
 For stability analysis, the flow is decomposed into the mean flow and 
 the disturbance according to
\begin{eqnarray}
 \tilde{u}_{i}(r,t)&=& U_{i}(r)+u_{i}(r,t),\\\label{5}
 \tilde{p}(r,t)&=&P(r)+ p(r,t). \label{eq6}
\end{eqnarray}

 We take  the   dimensional  base flow  for small suction and injection  (see \cite{3})
 \begin{eqnarray}
  U(y) &=& U^{*}(1-\frac{y^{2}}{h^{*2}})\label{eb} \\
  V&=& V^{*}_{\omega}\label{ebb}\\
  W&=&0.\label{ebbb}
 \end{eqnarray}
 By scaling  these velocities as above, we obtain with $h^{*}=\pm 1$ 
 the no-dimensional base flow  
 \begin{eqnarray}
  U(y) &=& (1-y^{2}) \label{eq*}\\
  V&=& 1\label{eq**}\\
  W&=&0.\label{eq***}
 \end{eqnarray}
 To obtain the stability equations for the spatial evolution of three-dimensional, 
 we take the
 dependent on time  disturbances 
\begin{eqnarray}
 (u(x, y, z, t);v(x, y, z, t);  w(x, y, z, t); p(x, y, z, t));
\end{eqnarray}
 which are  scaled in the same way as above.\\
 By using continuity, the pressure terms can be eliminated from Navier-Stokes equations.
  For such a mean profile (base flow), the divergence  of Navier-Stokes  equations
and continuity gives
  \begin{eqnarray}
   {\bigtriangledown}^{2}{p}&=& -2\frac{R_{e\omega}}{R_{e}}\frac{\partial{U}}{\partial{y}}
 \frac{\partial{v}}{\partial{x}}. \label{eq7}
 \end{eqnarray}
 
 The equations (\ref{eq2}) and (\ref{eq7})  after linearization give
 
 \begin{eqnarray}
  \frac{\partial{}}{\partial{t}} {\bigtriangledown}^{2}{v} + 
  {U} \frac{\partial{}}{\partial{x}}{\bigtriangledown}^{2}{v}+
   \frac{R_{e \omega}}{R_{e}}    
   \frac{\partial{}}{\partial{y}}{\bigtriangledown}^{2}{v} -
   \frac{d^{2}U}{{dy}^{2}}\frac{\partial{v}}{\partial{x}}
 &=& \frac{1}{R_{e}} {\bigtriangledown}^{4}{v}.\label{eq****}
\end{eqnarray}

The disturbances are taken 
to be periodic in the streamwise, spanwise directions and time,  
which allow us to assume solutions of the form
\begin{eqnarray}
 f(x,y,z,t)&=&\hat{f}(y)e^{i(\alpha x+\beta z-\omega t)}; \label{eq8}
\end{eqnarray}
where $f$ represents either one of the disturbances $u$, $v$, $w$ or $p$ and $\hat{f}$ 
the amplitude function, $k$, $\alpha= k_{x}=k\cos{\theta}$ and  
$\beta= k_{y}=k\sin{\theta}$ are the wave numbers, $\omega =\alpha c$ 
the pulsation of the wave.
With  $i^{2}=-1$,  $\theta= (\vec{k_{x}}, \vec{k})$, $c=c_{r}+ic_{i}$ wave velocity  
which is taken to be complex, $\alpha$ and $\beta$ are
real because of temporel stability analysis considered.\\
 Then with  the equation (\ref{eq8}), the equation (\ref{eq****}) becomes
 \begin{eqnarray}
  i\alpha \left[\left(U-c\right)\left(D^{2}-{k}^2  \right)
- {U''}\right]\hat{v} &=&-\frac{R_{e \omega}}{R_{e}}D\left(D^{2}-{k}^2  \right)\hat{v}\nonumber\\ 
 +\frac{1}{R_{e}}\left(D^{2} -{k}^{2} \right)^{2}\hat{v};\label{eq9}
\end{eqnarray}
where $D=\frac{d}{dy}$ ; with boundary conditions for all $(x,\pm1, z ,t)$
\begin{eqnarray}
    \hat{v}(\pm 1)&=&1 \nonumber\\
    \hat{v}'(\pm 1)&=&0. \label{eq10}
  \end{eqnarray}
  Taking
  \begin{eqnarray}
 v_{p}(x,y,z,t)&=& \hat{v}(y)e^{i(\alpha x+\beta z-\omega t)}-1,\label{eq11}
\end{eqnarray}
the equation (\ref{eq9}) and the boundary conditions  take the forms 
\begin{eqnarray}
  \left[  \left(  U- {\bf{i \alpha^{-1}\frac{R_{e \omega}}{R_{e}}D}}  \right)  \left( D^{2}-k^{2}\right)
  -U'' -i\alpha^{-1}\frac{1}{R_{e}}( D^{2}-k^{2})^{2}\right]\hat{v}_{p}= \nonumber\\
  c \left( D^{2}-k^{2}\right)\hat{v}_{p}\label{eq12},
  \end{eqnarray}
  \begin{eqnarray}
    \hat{v}_{p}(\pm 1)&=&0\nonumber\\
    \hat{v}_{p}'(\pm 1)&=&0. \label{eq13}
  \end{eqnarray}
 The equation (\ref{eq12})  is a flow equation modified by suction Reynolds number 
  ${R_{e \omega}}$ (or the speed of suction and injection),
  which we call modified Orr-Sommerfeld equation,  rewritten as an eigenvalue problem. 
\section{Stability analysis}
We  consider our three-dimensional disturbances. We use a temporal 
stability analysis as mentioned above. With $c$ complex as we have defined above,
when $c_{i}< 0$ we have stability, $c_{i}=0$ we have neutral stability and else we have instability.
We employ Matlab (Windows Version) in all our numerical computations to find the eigenvalues.
The Poiseuille parallel horizontal stationary porous  plates  flow with the basic profile  
\begin{eqnarray}
{\bf{U}} = \left(1-y^{2}, 1, 0 \right) 
\end{eqnarray}
 for $R_{e\omega}$ small ( i.e. small suction)  is considered  (see \cite{3}).
The eigenvalue problem (\ref{eq12}) is  solved numerically with a suitable boundary conditions.
The solutions are found in a layer bounded at $y=\pm1$ with ${\bf{U}}(\pm1)= (0, 1, 0)$.
The results of calculations are presented in the figures below.
We present the figures related to the eigenvalue problem (\ref{eq12}).

For all these  figures the black, red, green  and blue colors are respectively, 
for  $R_{e\omega}=0$, $R_{e\omega}=0.5$,  $R_{e\omega}=0.75,  $ $R_{e\omega}=1$ 
and the   yellow color is for $c_{i}=0$.

In each group of four figures, the first one is
Figure a), the second is Figure b), the third is Figure c) and the fourth is Figure d).

\begin{figure}[htbp]
 \begin{center}
 \includegraphics[width=6cm]{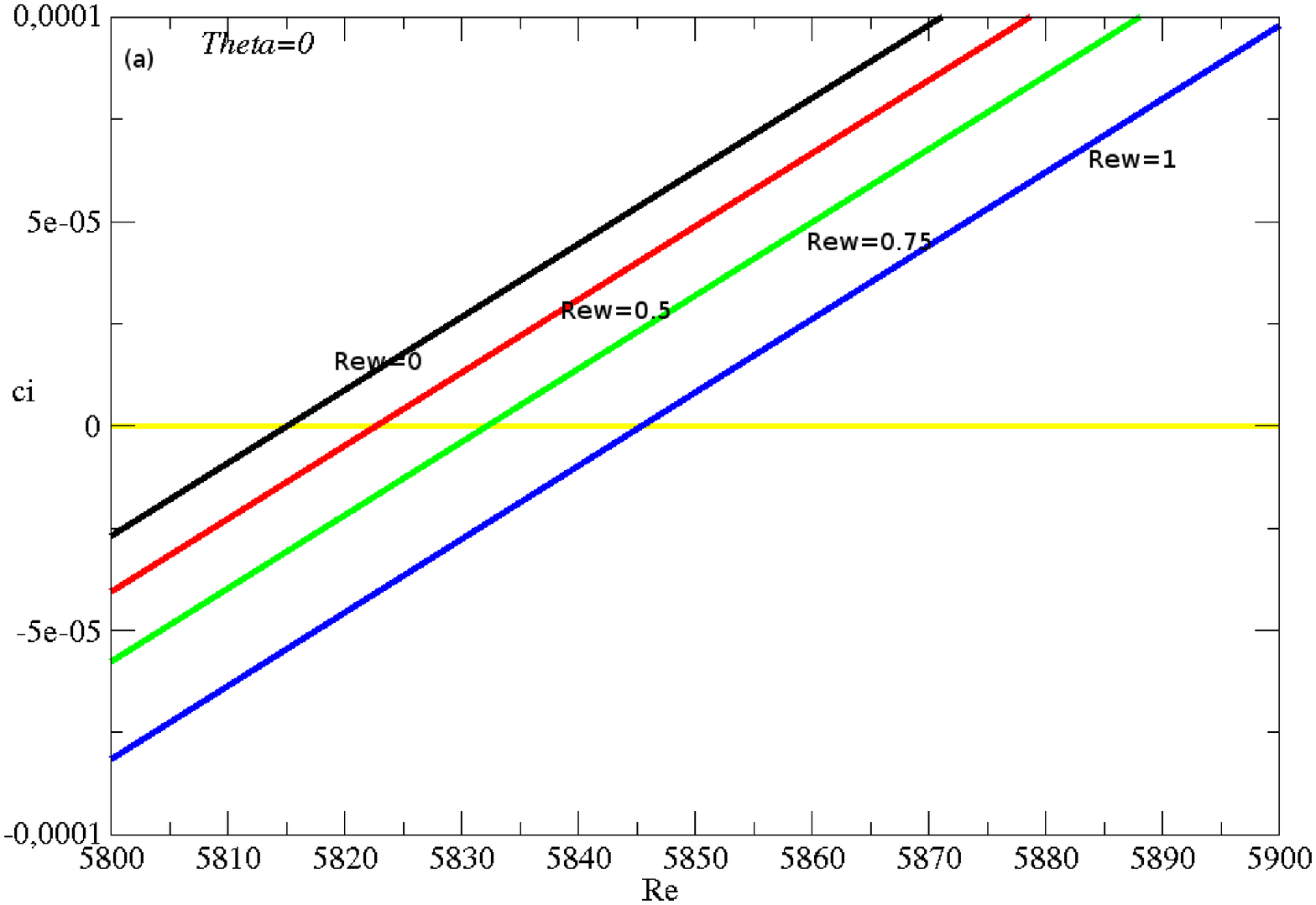}\includegraphics[width=6cm]{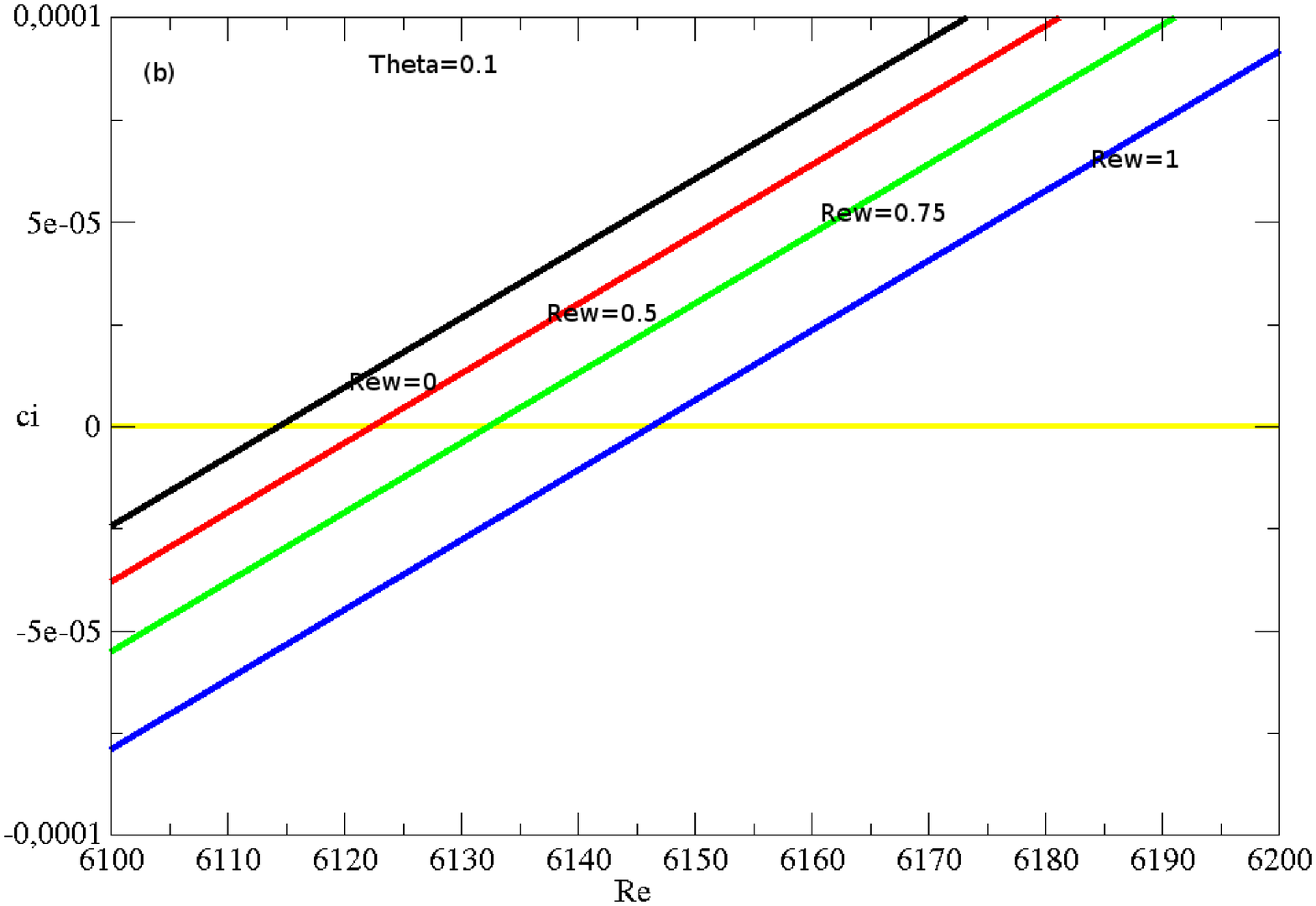}\\
 \includegraphics[width=6cm]{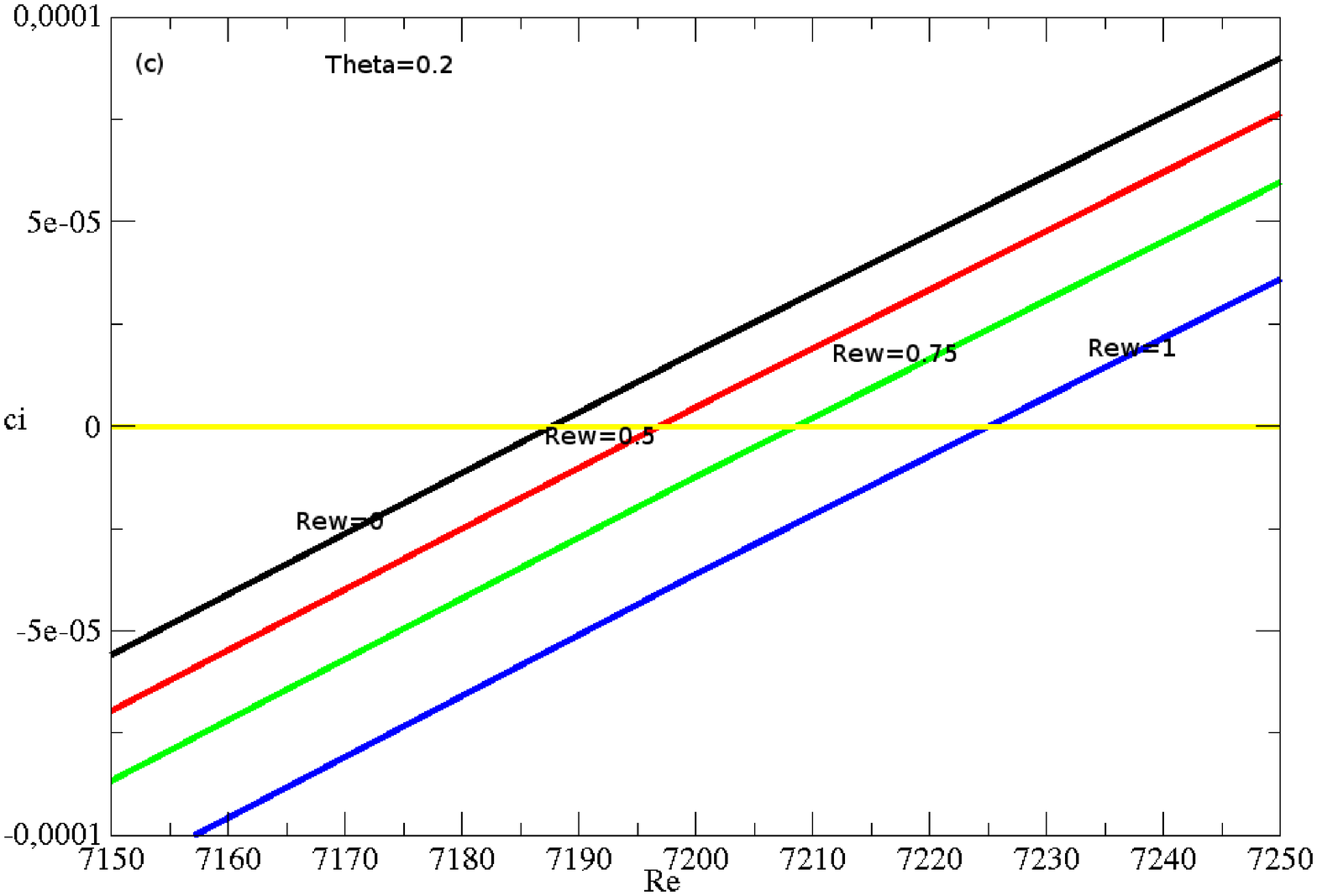}  \includegraphics[width=6cm]{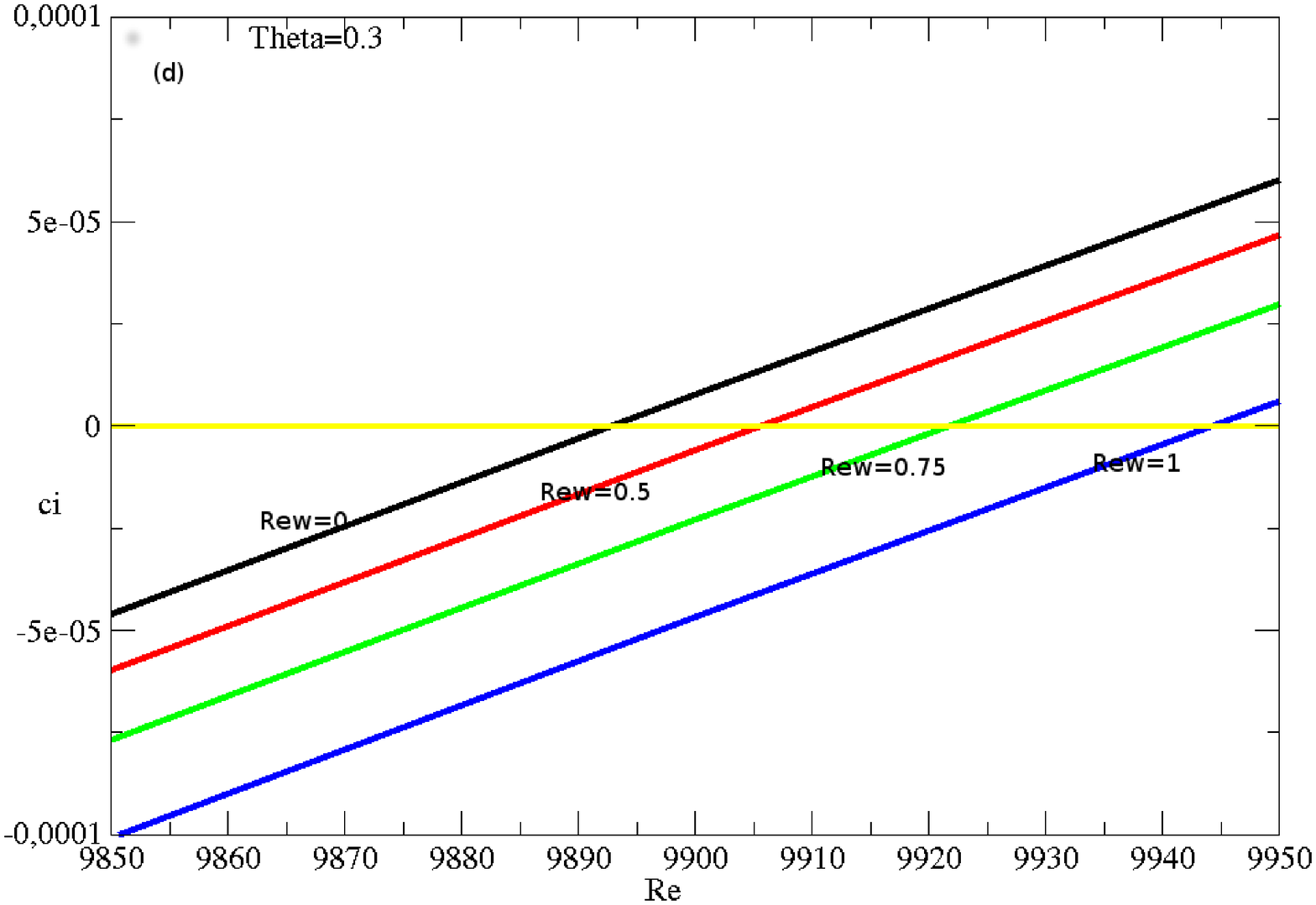}
 \end{center}
 \caption{Growth rate $c_{i}$ vs. Reynolds' number for $k=1$.}
 \label{fig:2}
 \end{figure}

For a fixed  $k=1$, we get figure (\ref{fig:2})  of $c_{i}$ vs. $R_{e}$ for sequential 
values of $\theta$.  figure $a$) for $\theta=0$, figure $b)$ for $\theta=0.1\pi$, 
figure $c)$ for $\theta=0.2\pi$ and  figure $d)$ for $\theta=0.3\pi$. 

\begin{figure}[htbp]
 \begin{center}
 \includegraphics[width=6cm]{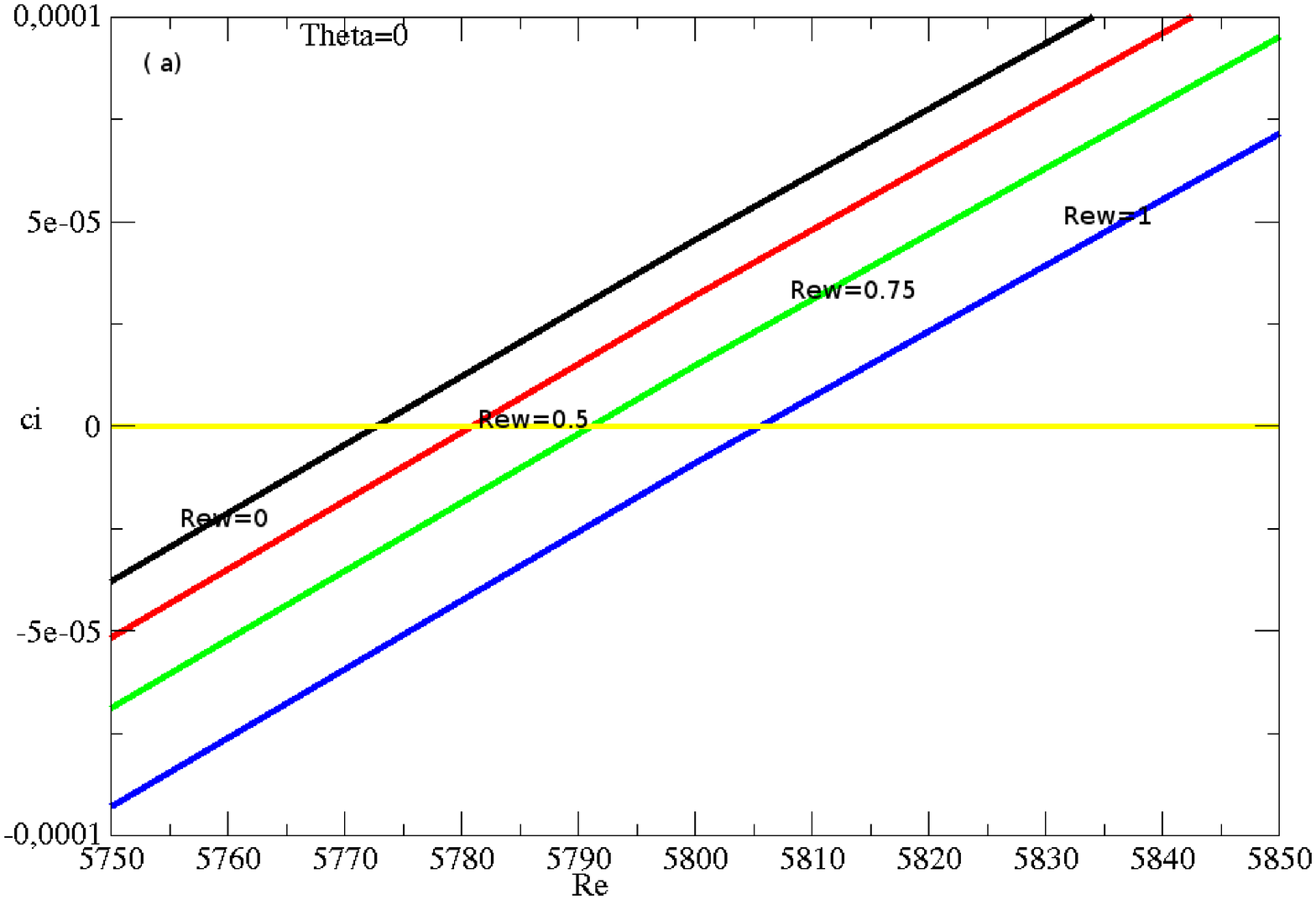}\includegraphics[width=6cm]{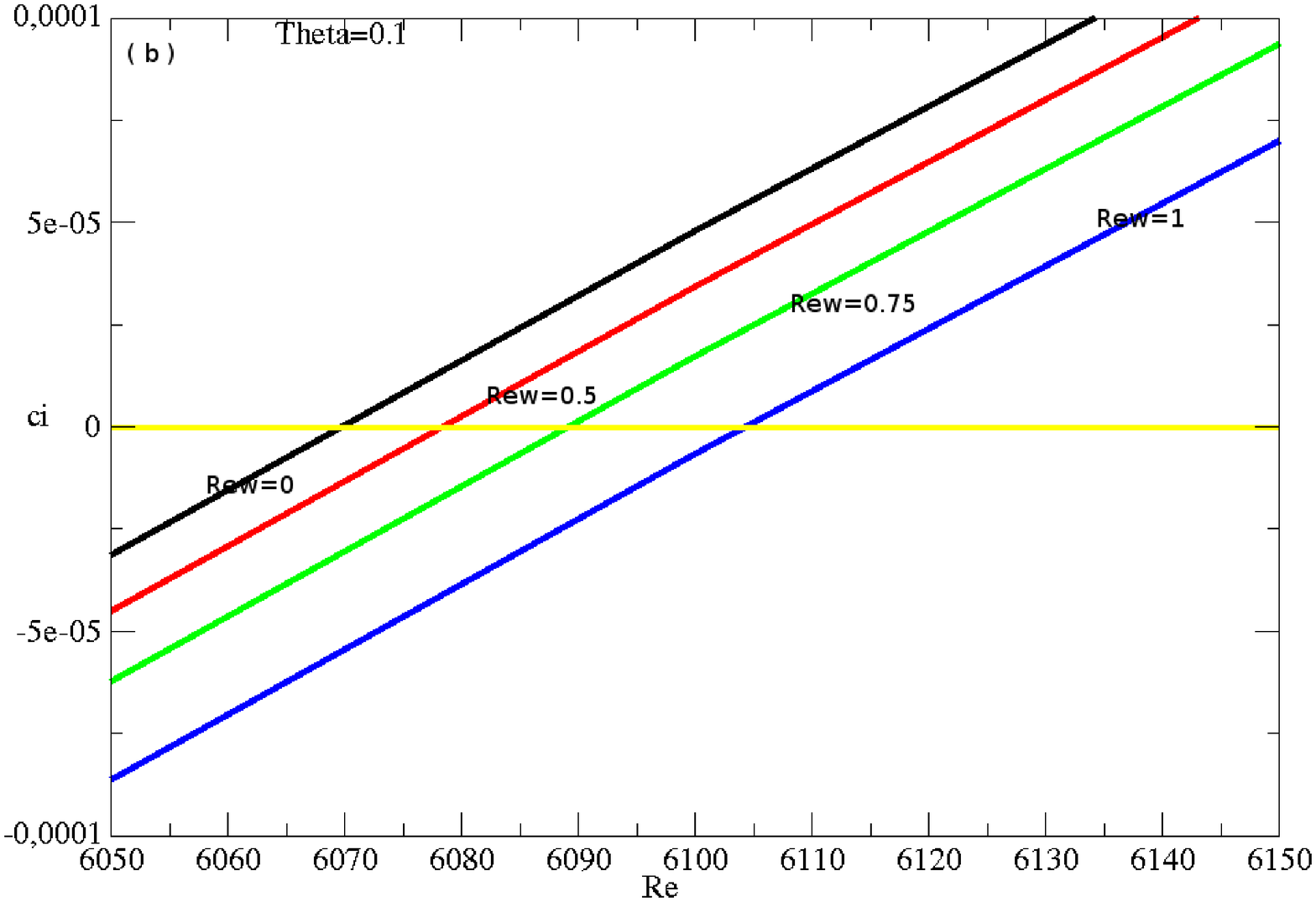}\\
 \includegraphics[width=6cm]{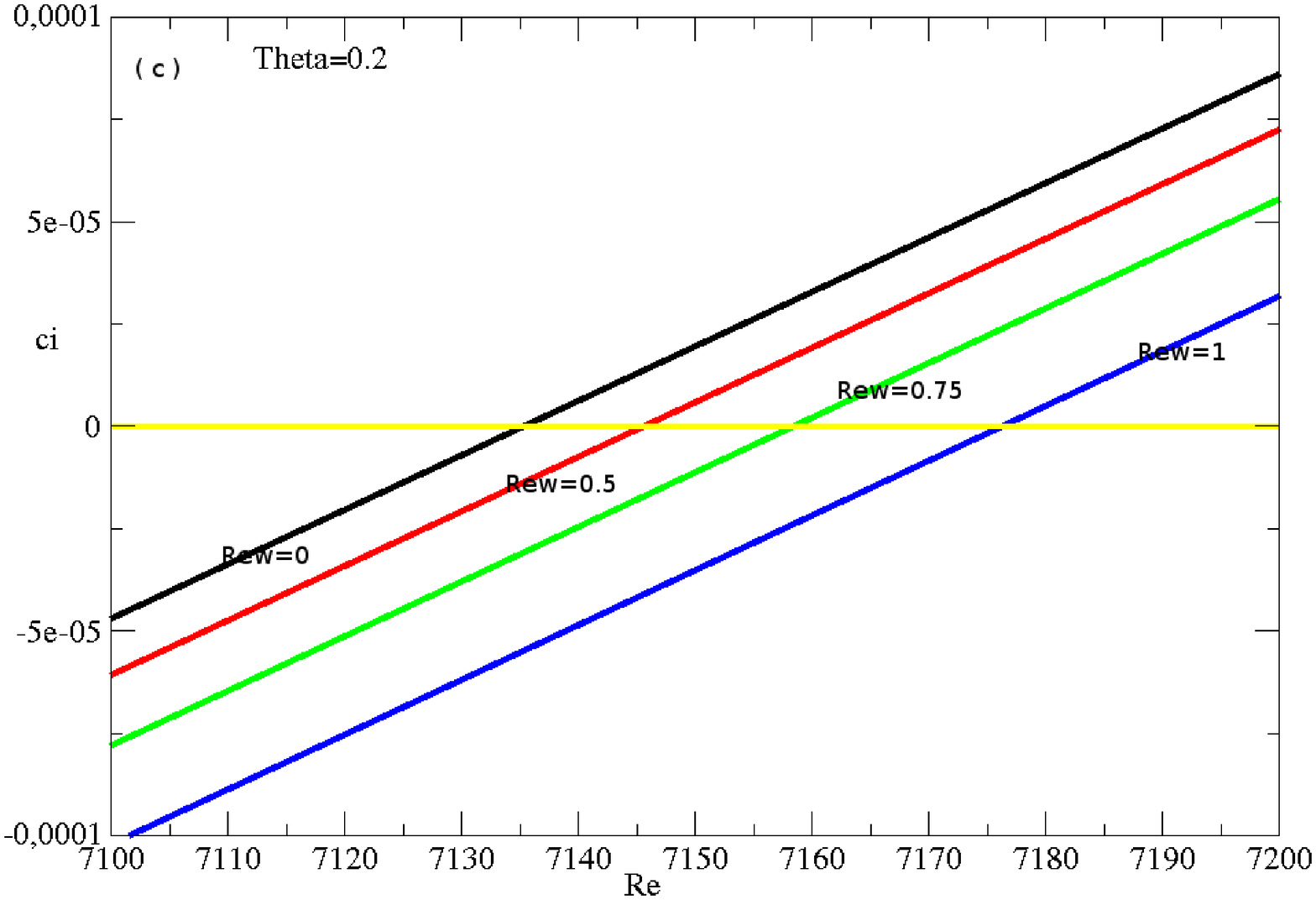}  \includegraphics[width=6cm]{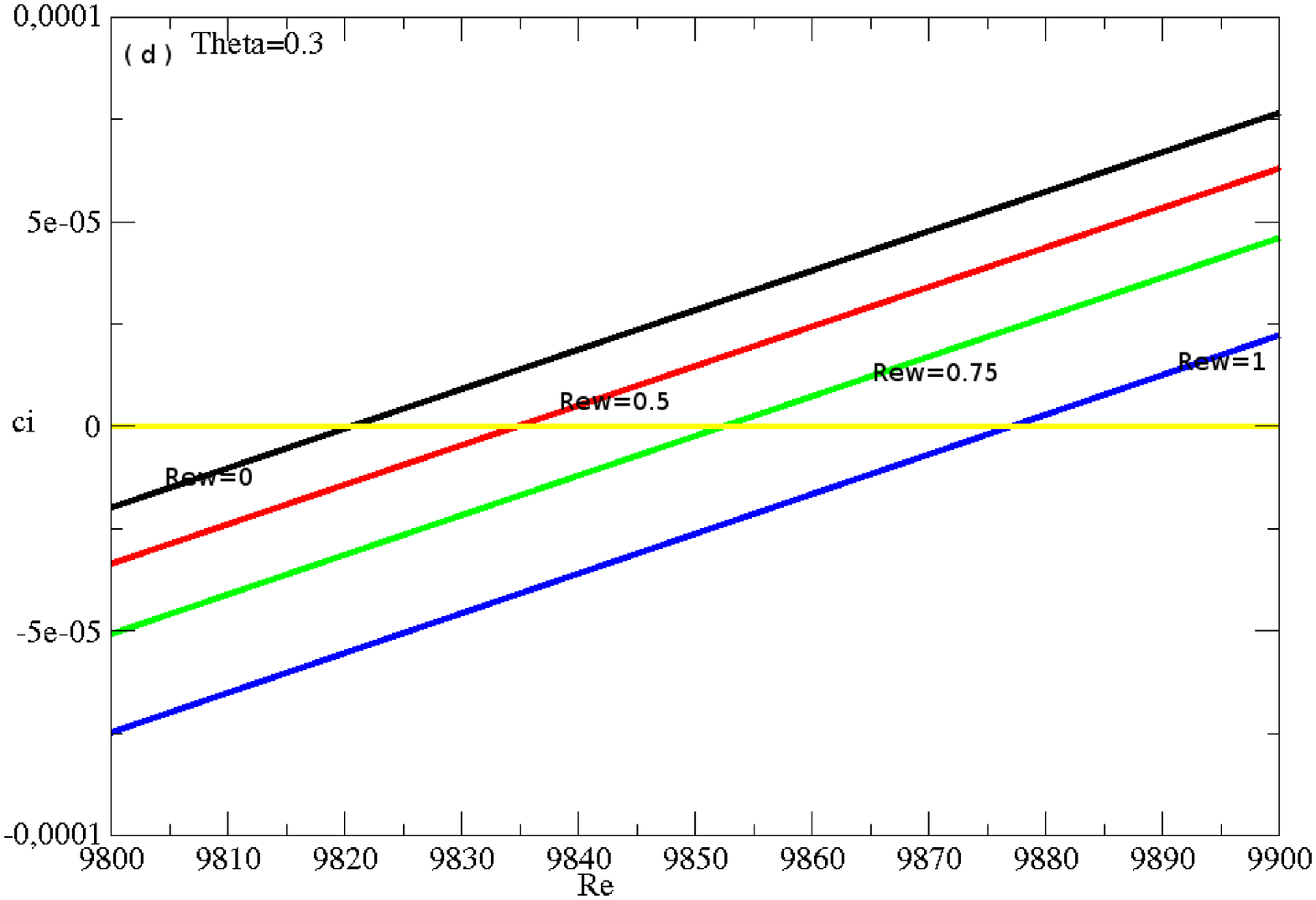}
 \end{center}
\caption{Growth rate $c_{i}$ vs. Reynolds' number for $k=1.02$.}
 \label{fig:3}
 \end{figure}

For a fixed  $k=1.02$, we get figure (\ref{fig:3})  of $ci$ vs. $R_{e}$ for sequential 
values of $\theta$.  figure $a$) for $\theta=0$, figure $b)$ for $\theta=0.1\pi$, 
figure $c)$ for $\theta=0.2\pi$ and  figure $d)$ for $\theta=0.3\pi$. 

\begin{figure}[htbp]
 \begin{center}
 \includegraphics[width=6cm]{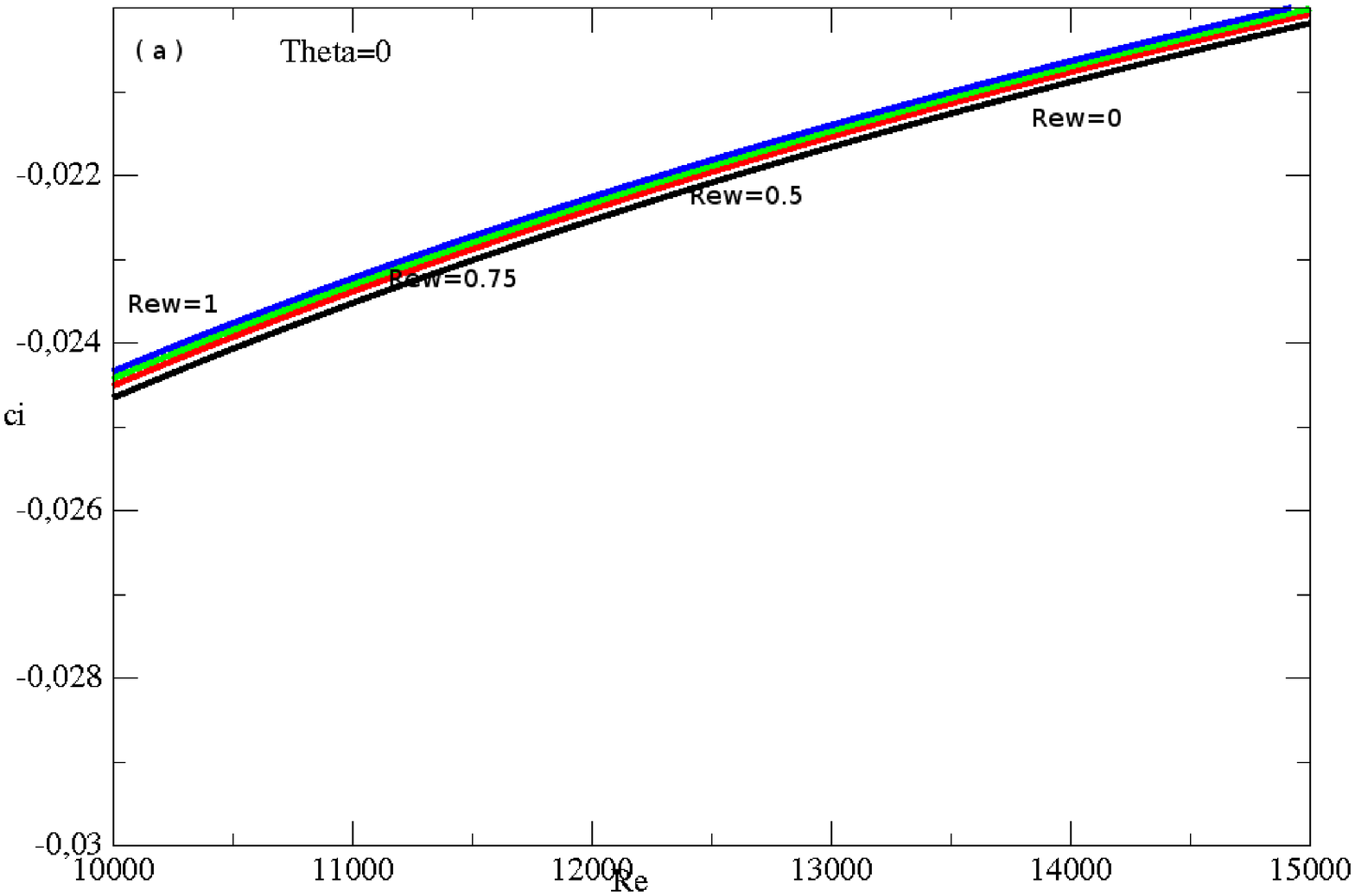}\includegraphics[width=6cm]{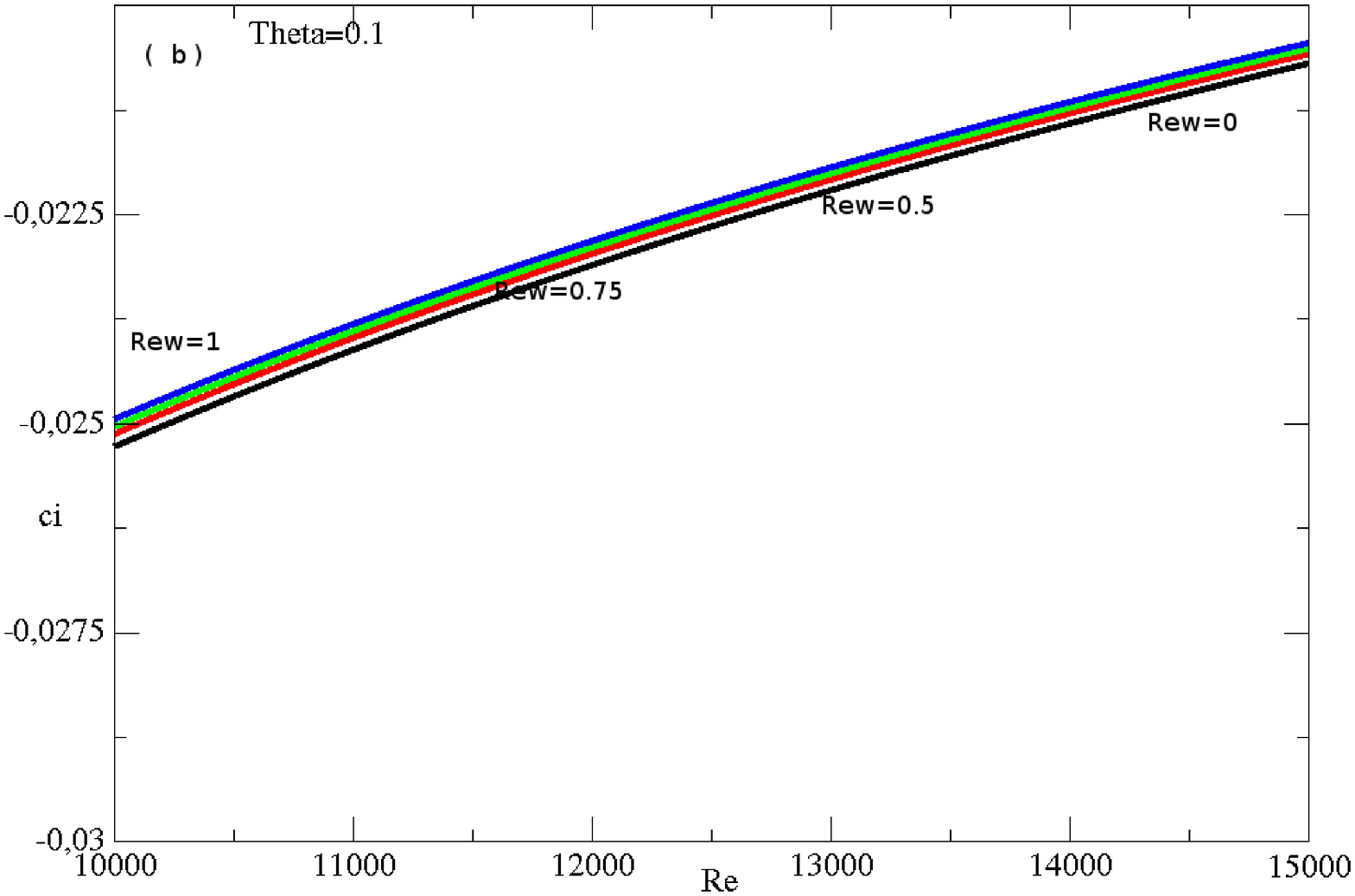}\\
 \includegraphics[width=6cm]{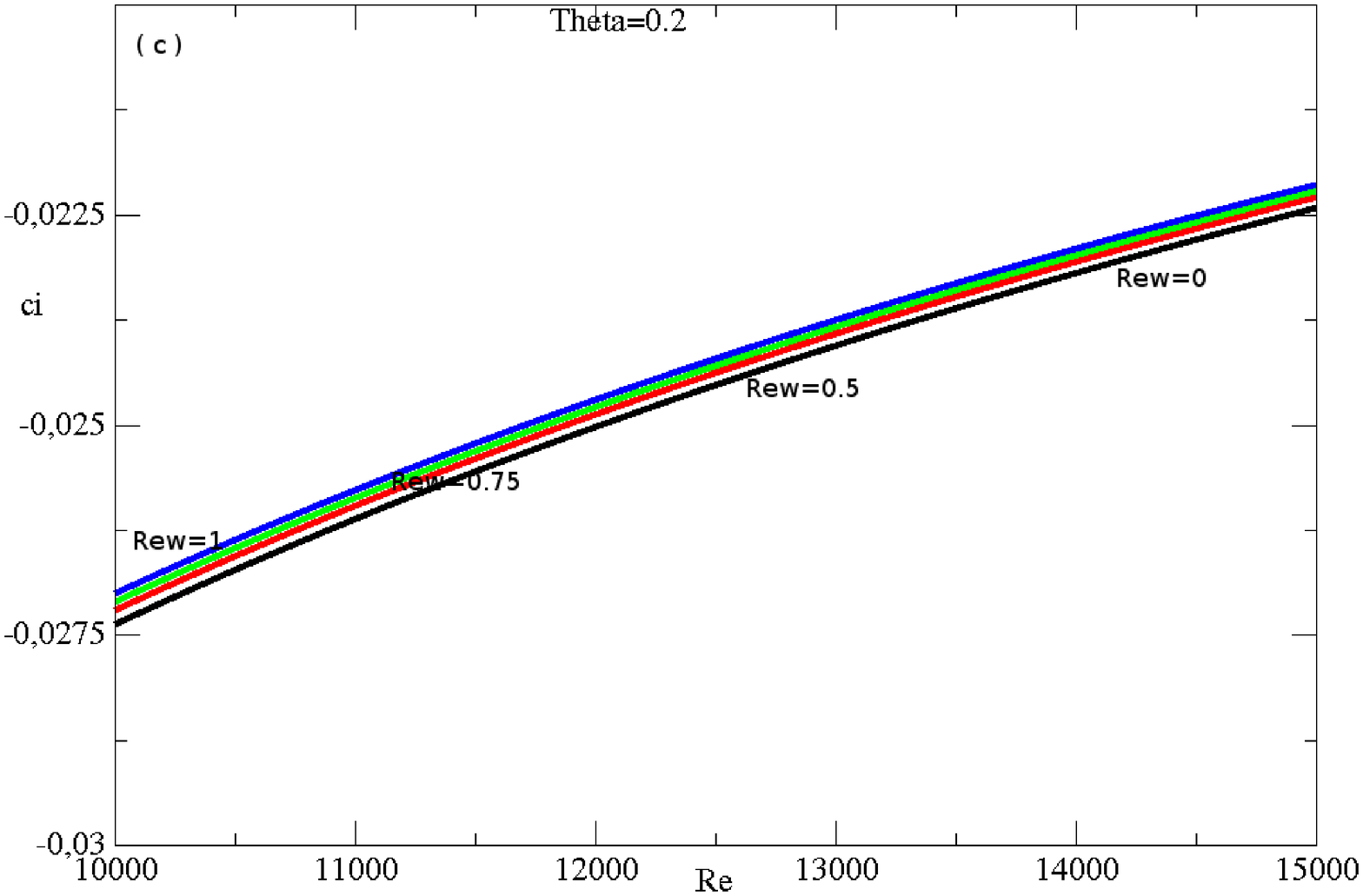}  \includegraphics[width=6cm]{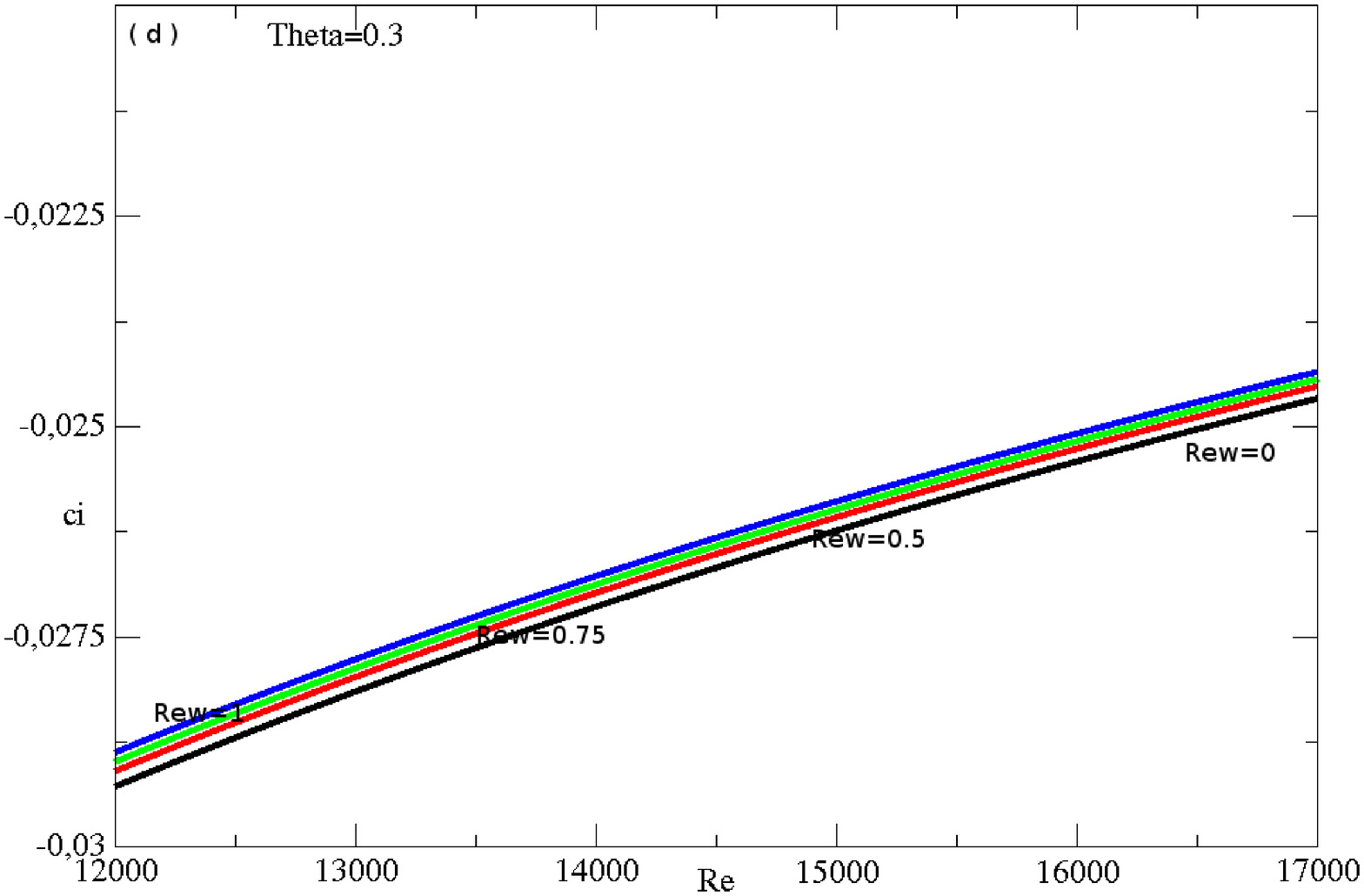}
 \end{center}
 \caption{Growth rate $c_{i}$ vs. Reynolds' number for $k=2$.}
 \label{fig:4}
 \end{figure}

For a fixed  $k=2$, we get figure (\ref{fig:4})  of $ci$ vs. $R_{e}$ for sequential 
values of $\theta$.  figure $a$) for $\theta=0$, figure $b)$ for $\theta=0.1\pi$, 
figure $c)$ for $\theta=0.2\pi$ and  figure $d)$ for $\theta=0.3\pi$.

\begin{figure}[htbp]
 \begin{center}
 \includegraphics[width=6cm]{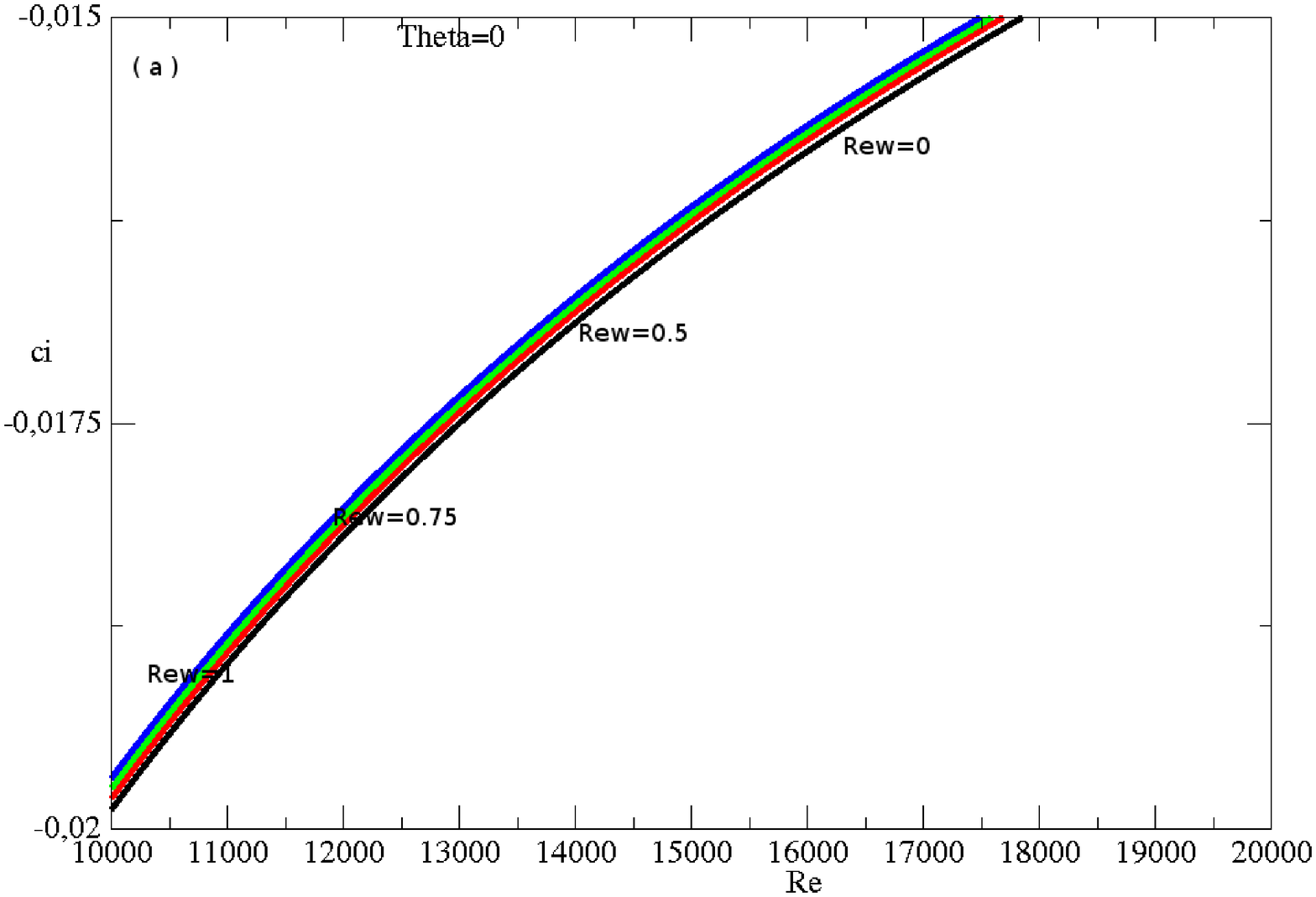}\includegraphics[width=6cm]{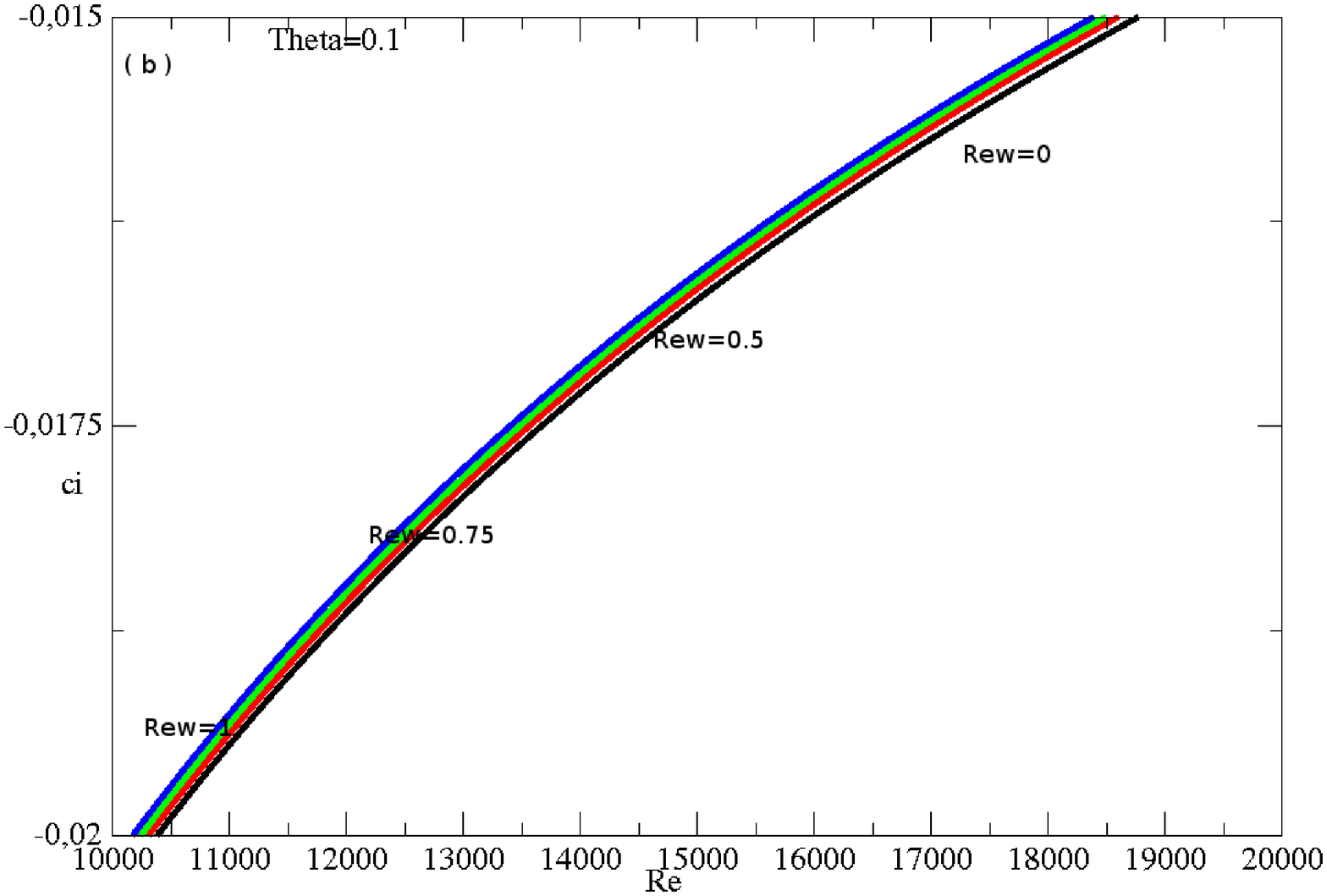}\\
 \includegraphics[width=6cm]{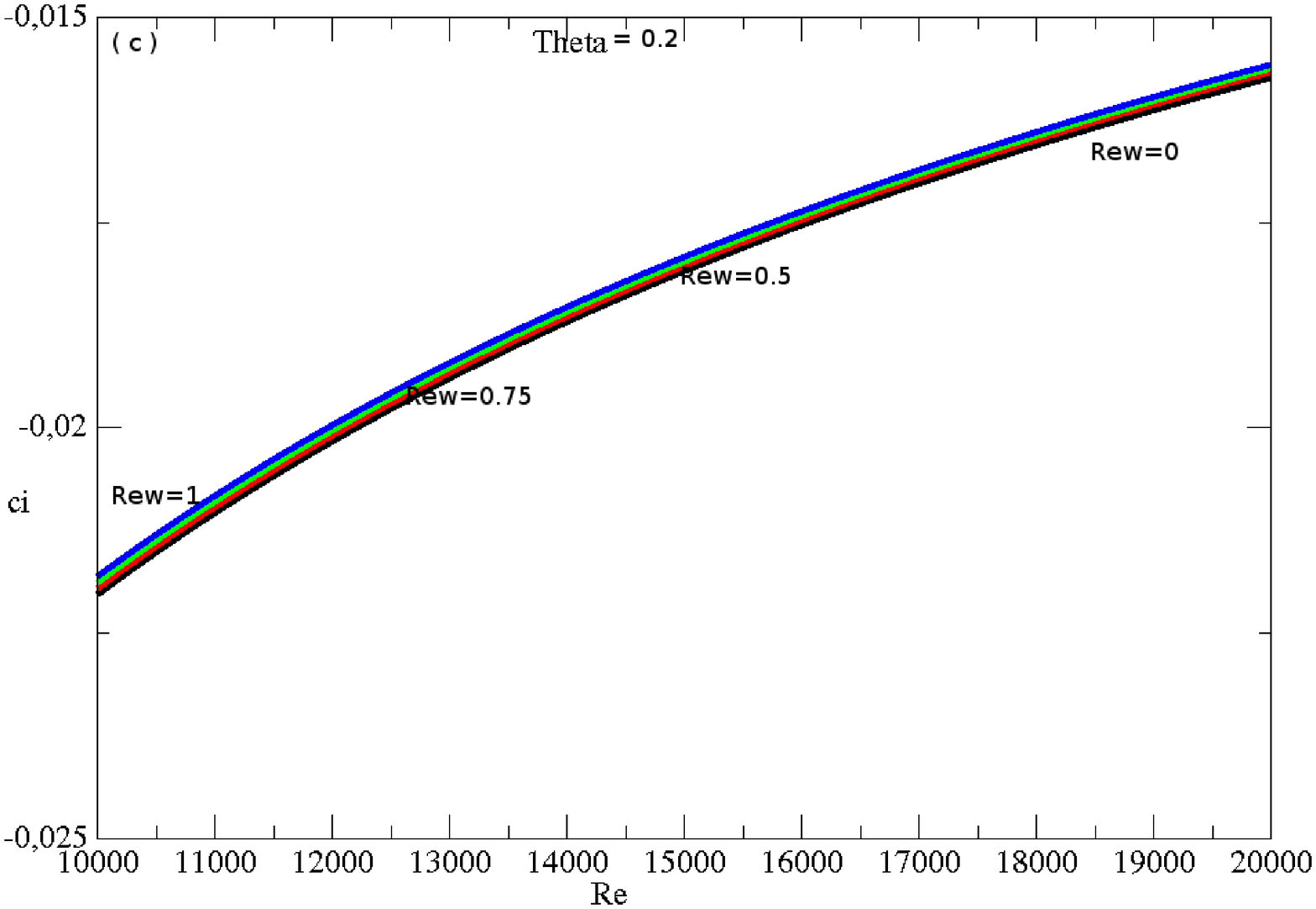}  \includegraphics[width=6cm]{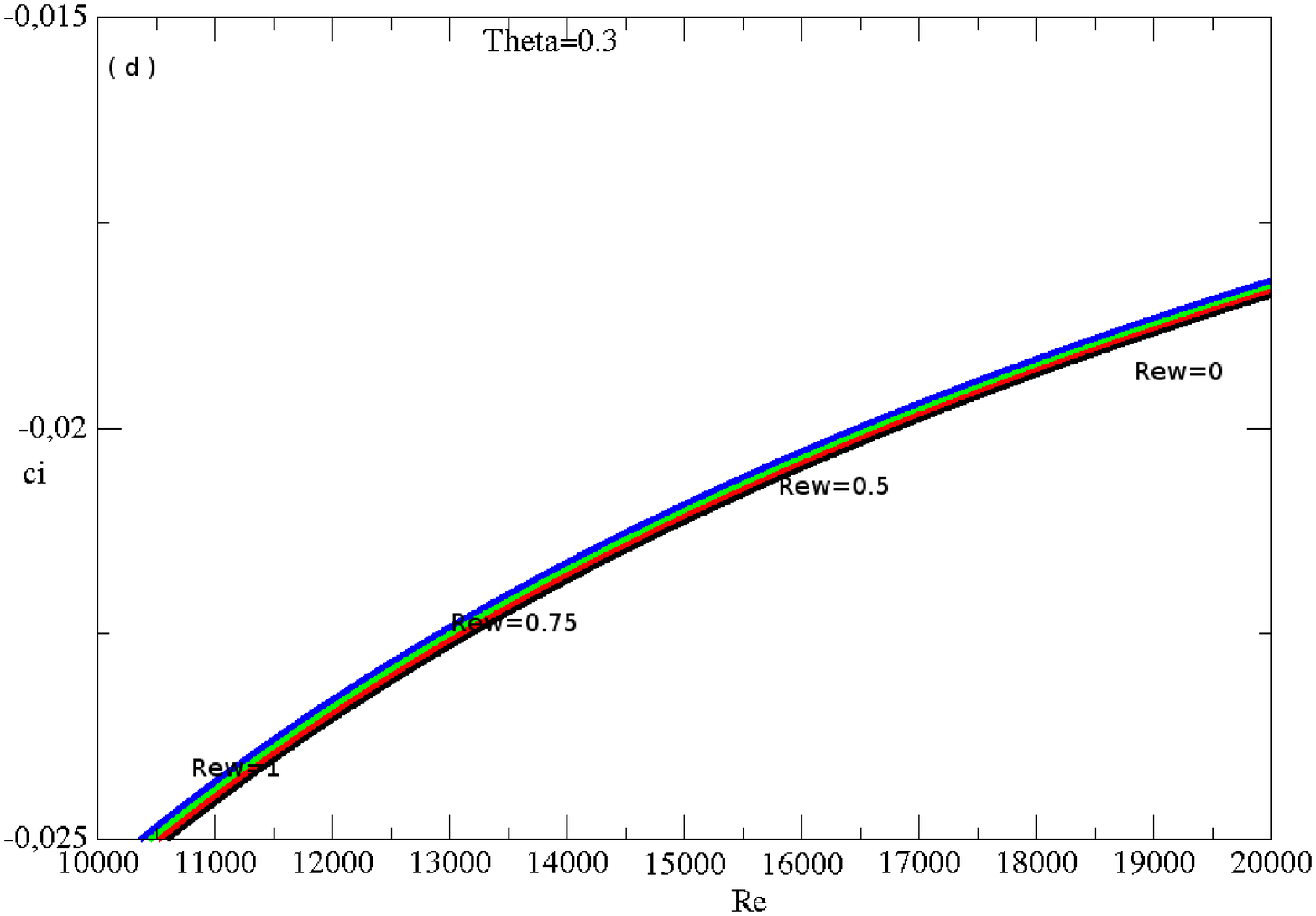}
 \end{center}
 \caption{Growth rate $c_{i}$ vs. Reynolds' number for $k=3$.}
 \label{fig:5}
 \end{figure}

For a fixed  $k=3$, we get figure (\ref{fig:5})  of $ci$ vs. $R_{e}$ for sequential 
values of $\theta$.  figure $a$) for $\theta=0$, figure $b)$ for $\theta=0.1\pi$, 
figure $c)$ for $\theta=0.2\pi$ and  figure $d)$ for $\theta=0.3\pi$. 

\begin{figure}[htbp]
 \begin{center}
 \includegraphics[width=7cm]{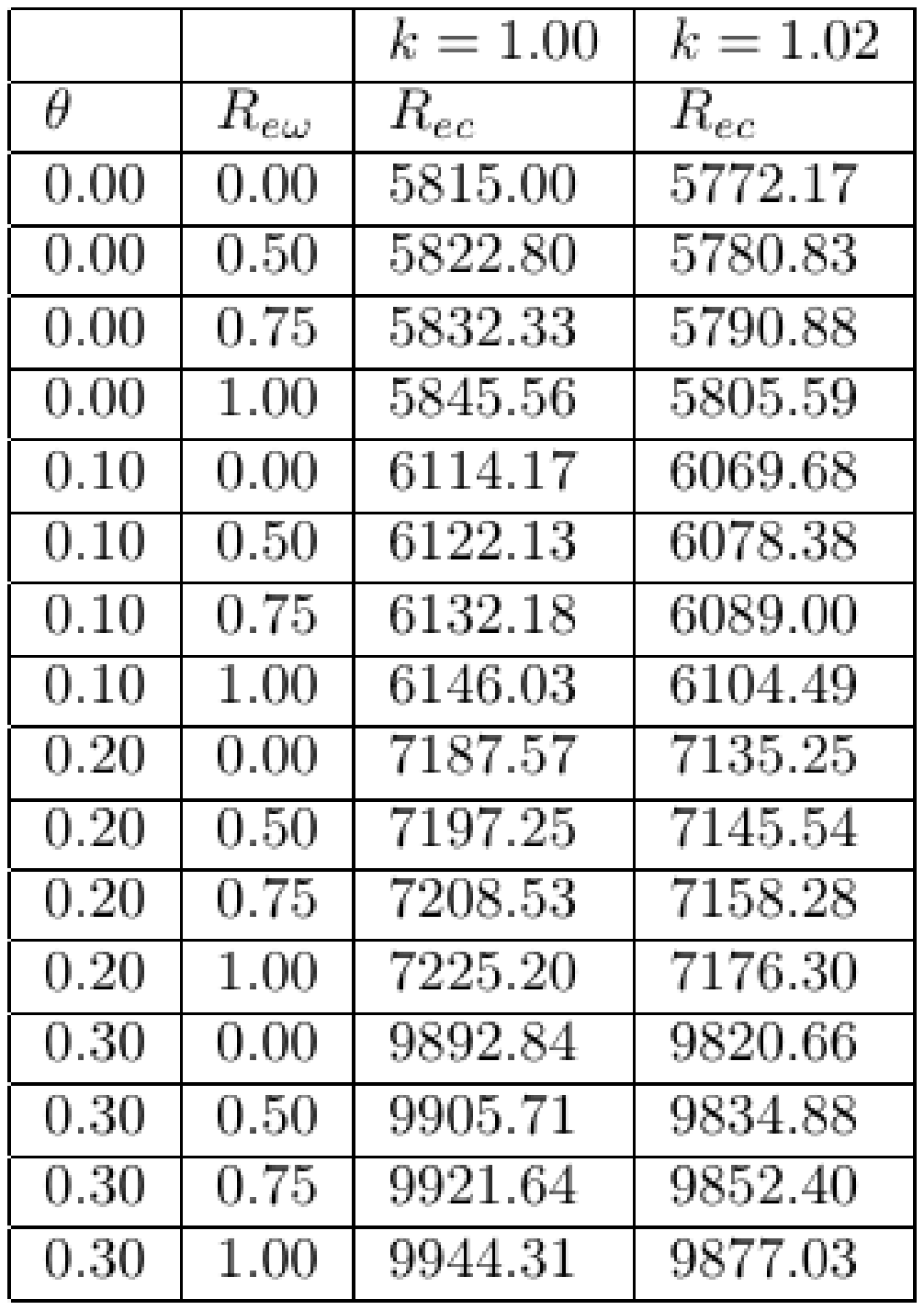} 
 \end{center}
 \caption{Tabular for critical Reynolds numbers} 
 \label{tabular:1}
 \end{figure}

Through the Figures (\ref{fig:2} and \ref{fig:3}), it is easy to see that the stability increases when 
 $R_{e\omega}$ increases for all $\theta$  because, 
for any   familly of curves of these figures the slope  fall when the suction  Reynolds' number 
increases. The  Reynolds' critical number $R_{ec}$ for which  we have the 
transition becomes important when $R_{e\omega}$ increases (see tabular (\ref{tabular:1}))
which  confirms that the wall small suction or injection have  a stabilizing effect on the
 viscous incompressible flow.
Because  of   $R_{e\omega}$  influences, we said that to normalize  with 
the characteristic velocity of suction   is necessary for  a perfect command 
of the field of stability.
In particular for ($R_{e\omega}=0,k=1.02 $, without suction, figure (\ref{fig:3}) black curve  ($a$) )
we find $R_{ec}=5772$ 
which corresponds exactly to the critical value given by classical linear theory  for a 
plane-Poiseuille flow without suction or injection.

 For $R_{e\omega} =1$, the equation  (\ref{eq12}) is identicaly for the equation $(2.28)$ found  by
 E. Niklas Davidsson and L. Hakan  Gustavsson in \cite{1}. For this value  of  $R_{e\omega}$ in the 
 cases of figures (\ref{fig:2} and \ref{fig:3})  the field of stability is more  
 greater  than the other small suction  Reynolds number for all $\theta$. 
  
Note that for great  $k$  the case of the Figures (\ref{fig:4}) and (\ref{fig:5})
there  is  stability without transition for all value of $\theta$ and $R_{e\omega}$ 
but when the small suction $R_{e\omega}$ increases the field of stability  decreases.
We can also say  that $k$ influences the stability of the flow i.e.
the growth of the wave number induces also the stability of the flow.
We therefore  conclude that in  two parallel horizontal stationary porous plates viscous and 
incompressible fluid flow, with small suction and small injection, the small suction 
Reynolds number stabilize  the  flow for small wave number but  the high wave number
 effect  is important than small suction Reynolds' number on the 
fluid flow stability.

 \section*{figures}
 
\newpage
 \section{Conclusion}
 In this paper, we have investigated  the  effect of small suction  
Reynolds number  on the   stability  of the fluid flow between two parallel horizontal 
stationary porous  plates . We have shown that the instability of the perturbed flow is 
governed by a remarkably equation named modified  Orr-Sommerfeld equation. 
We find also that the  normalization  of the wall-normal velocity with characteristic 
small suction (or small injection) velocity is important for a perfect command
of fluid flow stability analysis. We noticed  that for $k=1.02$ and $R_{e\omega}=0$
 we find $R_{ec}=5772$ which corresponds exactly to the critical value given by classical 
 linear theory  for a plane-Poiseuille flow without suction or injection. We  noticed also
 that  the  high  wave number stabilize more  than 
 the small suction Reynolds number.
 
\end{document}